\documentclass[11pt]{article}
\pdfoutput=1
\usepackage[centertags]{amsmath}
\usepackage[square, comma, sort&compress,numbers]{natbib}
\usepackage{array,multirow}
\numberwithin{equation}{section}
\usepackage{amssymb,mathtools}
\usepackage{amsfonts,amsthm,stmaryrd}
\usepackage{graphicx}
\usepackage{color,bm}
\usepackage[normalem]{ulem}

\usepackage{epsfig}
\usepackage{bbold}

\usepackage{wrapfig}
\usepackage{float}
\usepackage{soul}
\usepackage{color}
\usepackage{ulem}

\usepackage[dvipsnames]{xcolor}

\usepackage{tikz,pgf}
\usetikzlibrary{shapes}
\usetikzlibrary{calc}
\usetikzlibrary{decorations.pathmorphing}
\usetikzlibrary{decorations.pathreplacing,shapes.misc}
\usetikzlibrary{positioning}
\usetikzlibrary{arrows}
\usetikzlibrary{decorations.markings}
\usetikzlibrary{shadings}

\usetikzlibrary{intersections}

\def\sideremark#1{\ifvmode\leavevmode\fi\vadjust{\vbox to0pt{\vss
 \hbox to 0pt{\hskip\hsize\hskip1em
 \vbox{\hsize3cm\tiny\raggedright\pretolerance10000
  \noindent #1\hfill}\hss}\vbox to8pt{\vfil}\vss}}}

\newcommand{\be}{\begin{equation}}
\newcommand{\ee}{\end{equation}}
\newcommand{\ba}{\begin{eqnarray}}
\newcommand{\ea}{\end{eqnarray}}

\def\dps{\displaystyle}

\def\tr{{\rm Tr}}

\advance\textheight\topskip
\renewcommand{\tilde}{\widetilde}

\renewcommand{\simeq}{\cong}

\newcommand{\bref}[1]{\textbf{\ref{#1}}}

\newcommand{\binner}[2]{%
  {\langle}\kern-4.15pt{\langle}#1{,}\,#2{\rangle}\kern-4.15pt{\rangle}}

\newcommand{\half}{\mathchoice{%
    \ffrac{1}{2}}{\frac{1}{2}}{\frac{1}{2}}{\frac{1}{2}}}

\newcommand{\ffrac}[2]{\raisebox{.5pt}%
  {\footnotesize$\displaystyle\frac{#1}{#2}$}\kern1pt}

\numberwithin{equation}{section} \makeatletter
\@addtoreset{equation}{section}

\def\be{\begin{equation}}
\def\ee{\end{equation}}
\def\ba{\begin{array}}
\def\ea{\end{array}}

\def\dps{\displaystyle}
\def\tr{{\rm Tr}}

\newdimen\tableauside\tableauside=1.0ex
\newdimen\tableaurule\tableaurule=0.4pt
\newdimen\tableaustep
\def\phantomhrule#1{\hbox{\vbox to0pt{\hrule height\tableaurule
width#1\vss}}}
\def\phantomvrule#1{\vbox{\hbox to0pt{\vrule width\tableaurule
height#1\hss}}}
\def\sqr{\vbox{%
  \phantomhrule\tableaustep

\hbox{\phantomvrule\tableaustep\kern\tableaustep\phantomvrule\tableaustep}%
  \hbox{\vbox{\phantomhrule\tableauside}\kern-\tableaurule}}}
\def\squares#1{\hbox{\count0=#1\noindent\loop\sqr
  \advance\count0 by-1 \ifnum\count0>0\repeat}}
\def\tableau#1{\vcenter{\offinterlineskip
  \tableaustep=\tableauside\advance\tableaustep by-\tableaurule
  \kern\normallineskip\hbox
    {\kern\normallineskip\vbox
      {\gettableau#1 0 }%
     \kern\normallineskip\kern\tableaurule}%
  \kern\normallineskip\kern\tableaurule}}
\def\gettableau#1 {\ifnum#1=0\let\next=\null\else
  \squares{#1}\let\next=\gettableau\fi\next}

\def\cA{\mathcal{A}}
\def\cB{\mathcal{B}}

\def\cD{\mathcal{D}}

\def\cF{\mathcal{F}}

\def\cH{\mathcal{H}}

\def\cK{\mathcal{K}}

\def\cM{\mathcal{M}}

\def\cR{\mathcal{R}}

\def\cV{\mathcal{V}}

\def\cZ{\mathcal{Z}}

\def\bA{\bold A}
\def\bB{\bold B}
\def\bF{\bold F}
\def\bW{\bold W}

\def\bZ{\bold Z}

\def\bPsi{\bold \Psi}
\def\bchi{{\boldsymbol \chi}}
\def\bxi{{\boldsymbol \xi}}

\def\bchi{{\boldsymbol \chi}}

\numberwithin{equation}{section} \makeatletter
\@addtoreset{equation}{section}

\newcommand{\idm}{\mathbb{I}_N }		
\newcommand{\bT}{\mathbf{T} }	

\def\ads2{\text{AdS}_{2}}
\def\cft1{\text{CFT}_{1}}
\def\hs{$hs[\lambda]$}

\def\be{\begin{equation}}
\def\ee{\end{equation}}
\def\ba{\begin{array}}
\def\ea{\end{array}}

\def\dps{\displaystyle}

\def\ba{\begin{array}}
\def\ea{\end{array}}

\def\dps{\displaystyle}

\def\hs{\mathfrak{hs}[\lambda]}

\numberwithin{equation}{section} 
\numberwithin{figure}{section} 
\numberwithin{table}{section} 

\usepackage{sectsty} 
\allsectionsfont{\centering } 

\def\XXint#1#2#3{{\setbox0=\hbox{$#1{#2#3}{\int}$}
     \vcenter{\hbox{$#2#3$}}\kern-.5\wd0}}

\def \Tr{\mbox{Tr\,}}

\def \tr{\mbox{tr\,}}
\def \rd{{\rm{d}}}

\usepackage{jheppub}

\makeatletter
\def\@fpheader{\vspace{-.1cm}}
\makeatother

\title{\centering{Color decorations of Jackiw-Teitelboim  
gravity}}

\author[a]{Konstantin\ Alkalaev,} 
\author[b]{Euihun\ Joung,}
\author[c,d,e]{Junggi\ Yoon}

\affiliation[a]{I.E. Tamm Department of Theoretical Physics,\\ P.N. Lebedev Physical Institute, \\Leninsky ave. 53, 119991 Moscow, Russia}
\affiliation[b]{Department of Physics, Kyung Hee University,\\ 26 Kyungheedae-ro Dongdaemun-gu, Seoul 02447, Korea}
\affiliation[c]{Asia Pacific Center for Theoretical Physics,\\77 Cheongam-ro, Nam-gu, Pohang-si, Gyeongsangbuk-do, 37673, Korea}
\affiliation[d]{Department of Physics, POSTECH\\ 77 Cheongam-ro, Nam-gu, Pohang-si, Gyeongsangbuk-do, 37673, Korea}
\affiliation[e]{School of Physics, Korea Institute for Advanced Study\\
85 Hoegiro Dongdaemun-gu, Seoul 02455, Korea}

\emailAdd{alkalaev@lpi.ru}
\emailAdd{euihun.joung@khu.ac.kr}
\emailAdd{junggi.yoon@apctp.org}

\abstract{We introduce the colored version of Jackiw-Teitelboim (JT) gravity which is the two-dimensional dilaton gravity model with matrix-valued fields.  It is straightforwardly formulated in terms of BF action with $su(N,N)$ gauge algebra so that the standard JT gravity is embedded as $su(1,1) \subset su(N,N)$ subsector. We also elaborate on the respective metric formulation which is shown to involve the JT fields plus $su(N)$ non-Abelian  fields as well as $su(N)$-matrix valued metric and dilaton fields. Their interactions are governed by minimal couplings and potential terms of cubic and quartic orders  involving derivatives.
}

\begin{document}

\maketitle
\flushbottom

\section{Introduction}

The AdS$_2$/SYK correspondence~\cite{KitaevTalks,kitaevfirsttalk,Polchinski:2016xgd,Jevicki:2016bwu,Maldacena:2016hyu} identifies JT gravity \cite{Teitelboim:1983ux,Jackiw:1984je} as the bulk dual of the soft mode of the SYK model~\cite{KitaevTalks,kitaevfirsttalk,Maldacena:2016hyu,Maldacena:2016upp}. The issue of finding a complete dual  theory in AdS$_2$ containing JT gravity is basically open  (see e.g. a discussion in \cite{Gross:2017vhb,Gross:2017aos,Rosenhaus:2018dtp}) that encourages the search for extended JT-type models which could incorporate an infinite SYK spectrum~\cite{Das:2017pif,Das:2017hrt}. 

Note that the original JT gravity can be extended by adding more fields following  two different tracks using either  metric or frame (BF) formulations. Introducing a metric explicitly can be more useful in the context of finding exact solutions in dilaton gravity models in the presence matter  or gauge fields, see e.g.~\cite{Castro:2008ms,Gaikwad:2018dfc,Gonzalez:2018enk,Lala:2019inz,Lala:2020lge}. Within the BF formulation,  which is essentially the Cartan approach  to gravity, all extensions of the original theory are basically boiled down to extending the gauge algebra. A direct product  $sl(2, \mathbb{R})\times \cK$, where $\cK$ is some Lie algebra provides Yang-Mills type  extensions, see e.g. \cite{Gonzalez:2018enk}. Another way is to embed $sl(2, \mathbb{R}) \subset \cH$, where $\cH$ is some (in)finite-dimensional Lie algebra. For example, the extension of the gauge algebra   $sl(2, \mathbb{R})$ to higher-rank gauge algebras $sl(M,\mathbb{R})$  reveals higher-spin  JT gravity models with finite spectra of higher-rank fields \cite{Alkalaev:2013fsa,Grumiller:2013swa,Gonzalez:2018enk}. Moreover,  there are higher-spin JT models with infinite number of fields \cite{Alkalaev:2014qpa,Alkalaev:2019xuv,Alkalaev:2020kut,Vasiliev:1995sv} which are based on the infinite-dimensional extension of the gauge algebra $sl(2, \mathbb{R})$, known as the algebra $sl[\lambda]$
or $hs[\lambda]$, with a real parameter $\lambda$ \cite{Feigin,Bergshoeff:1989ns,Vasiliev:1989re}.\footnote{Such higher-spin models could be relevant in understanding AdS$_2$/SYK duality from the higher-spin perspective, see e.g. some suggesting avenues for discussion in \cite{Jevicki:2016bwu,Mezei:2017kmw,Gross:2017vhb,Gonzalez:2018enk,Peng:2018zap,Alkalaev:2019xuv,Alkalaev:2020kut,Alkalaev:2021zda}. Other possible modifications of the original JT gravity include various limits like Newton–Cartan and Carrollian versions \cite{Grumiller:2020elf,Gomis:2020wxp} which could be  extended using the respective higher-spin algebras recently discussed in \cite{Campoleoni:2021blr}.}

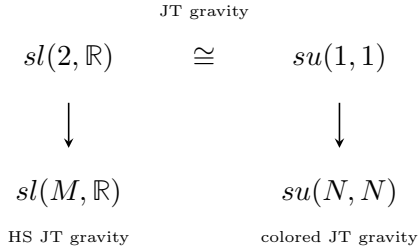
\begin{figure}[t]
\centering

\begin{tikzpicture}[line width=0.6pt,scale=0.60]

\draw (0,0) node {$sl(2, \mathbb{R})$};
\draw [-stealth](0,-1) -- (0,-2);
\draw (0,-3) node {$sl(M, \mathbb{R})$};
\draw (0,-4) node {$\text{\tiny{HS JT gravity}}$};

\draw (6,0) node {$su(1,1)$};
\draw [-stealth](6,-1) -- (6,-2);
\draw (6,-3) node {$su(N,N)$};
\draw (6,-4) node {$\text{\tiny{colored JT gravity}}$};

\draw (3,0) node {$\cong$};
\draw (3,1) node {$\text{\tiny{JT gravity}}$};

\end{tikzpicture}

\caption{Two possible tracks to extend  original AdS$_2$ global isometry algebra and respective extensions of JT gravity which in both cases is embedded as a subsector.} 
\label{fig1}
\end{figure}

In this paper we take the second route and introduce a colored version  of JT gravity which can be obtained by promoting the algebra  $sl(2, \mathbb{R}) \cong su(1,1)$   to the higher-rank algebra $su(N,N)$. The corresponding  BF theory naturally extends JT gravity by adding matrix-valued fields and can be viewed as the dilaton gravity carrying Chan-Paton color charges, see Fig. \bref{fig1}. Such a choice of the gauge algebra is naturally inherited  from 3d colored AdS (higher-spin) gravity \cite{Gwak:2015vfb,Joung:2017hsi,Gwak:2015jdo} and $3d$ colored Poincare gravity \cite{Gomis:2021irw} as well as their non-relativistic limits \cite{Joung:2018frr}.   

The paper is organized as follows. In Section \bref{sec:JT} we shortly review both BF and metric formulations of JT gravity.  Section \bref{sec:coloredJT} introduces the colored JT gravity as BF theory with $su(N,N)$ gauge algebra. Here, we address the gauging procedure and describe the resulting spectrum of the theory. In Section \bref{sec:2nd_order} we develop the metric formulation by solving the constraints for the singlet auxiliary fields arising in the JT sector of the theory. However, the resulting action still contains matrix auxiliary fields subject to complicated matrix constraints that can be solved only perturbatively. Nonetheless, it is still possible to find the general form of the second order colored gravity action that can be split into manifest JT gravity and $su(N)$ BF actions along with colored field contributions defined by implicit second order derivative  terms and explicit algebraic terms. Conclusions and future perspectives including preliminary notes on the colored AdS backgrounds and the color symmetry breaking are discussed in Section \bref{sec: discussions}.  Appendix \bref{app:BF} contains various technical details.

\section{JT gravity as $su(1,1)$ BF theory}
\label{sec:JT}

In this section we briefly  recapitulate the JT model in the metric form  
\cite{Teitelboim:1983ux,Jackiw:1984je} and its BF reformulation in terms $o(2,1)$ connections \cite{Fukuyama:1985gg,Chamseddine:1989wn,Isler:1989hq} for the purpose of setting the convention. Let $J_A$ with $A,B,...=0',0,1$ denote $o(2,1)$ basis elements. The commutation relations are given by\footnote{The $o(2,1)$-covariant Levi-Civita symbol $\epsilon_{ABC}$ is defined by $\epsilon_{0'01}=1$. The $o(2,1)$-invariant  metric is $\eta_{AB}={\rm diag}(--+)$. Note that the dS$_2$ case would correspond to the choice $\eta_{AB}={\rm diag}(+-+)$. The Lorentz subalgebra  $o(1,1)\subset o(2,1)$ has the tangent indices $a,b,... = 0,1$. The $o(1,1)$-covariant Levi-Civita symbol $\epsilon_{ab}$ is defined by 
$\epsilon_{ab} = \epsilon_{0'ab}$ and $\epsilon^{ab} = -\epsilon^{0'ab}$
so that $\epsilon_{01}=1$,
and it satisfies the identity $\epsilon_{ab}\epsilon^{mn} = -(\delta_a^n\delta_b^m -\delta_a^m\delta_b^n)$. The $o(1,1)$-invariant  metric is $\eta_{ab}={\rm diag}(-+)$. The Levi-Civita symbol $\epsilon_{\mu\nu}$ and Lorentz tensor $\eta_{\mu\nu}$ with the world indices $\mu,\nu,... = 0,1$
 are defined by the same relations. We omit the wedge product $\wedge$ symbol in  exterior products: $A_{(p)}\wedge B_{(s)} \equiv A_{(p)} B_{(s)}$.  } 
\be
\label{com}
[J_A,J_B]=2\epsilon_{ABC}\,\eta^{CD}\,J_D\,.
\end{equation}
The fields are differential 0-form $\Phi=\Phi^A\,J_A$ and 1-form $\Omega=\rd x^\mu\,\Omega_\mu^A\, J_A$ taking values in the $o(2,1)$ adjoint representation. The BF formulation of  JT gravity on a two-dimensional manifold $\cM_2$ is given by the following action \cite{Fukuyama:1985gg,Chamseddine:1989wn,Isler:1989hq}
\be
\label{BF}
S_{_{JT}}[\Omega, \Phi]=\kappa\int_{\cM_2}  R_A \Phi^A \,,
\ee
where $\kappa$ is a dimensionless coupling constant and  2-form curvature reads $R_A=\rd\Omega_A+ \epsilon_{ABC}\, \Omega^B \Omega^C$.

The equations of motion that follow from the BF action \eqref{BF} have the form of the covariant constancy condition and the zero-curvature condition,
\be
\label{BF_eom}
\rd\Phi_{A}+2\epsilon_{ABC}\,\Omega^B\, \Phi^C  = 0\;,
\qquad
\rd\Omega^A+\epsilon^{ABC}\,\Omega_B \Omega_C=0\;.
\ee 

Now the $o(2,1)$ basis elements $J_A$ are split as translations $P_a = (J_{0}, J_1)$ and Lorentz rotation $L = J_{0'}$ with  the commutation relations:  
$\left[ L,P_a \right] =2\epsilon_{ab}P^b$, and $\left[ P_a,P_b \right] = -2\epsilon_{ab}\,L$. This allows representing  the fields and the curvature in the Lorentz basis as $\Omega^A = (\omega, e^a)$, $R^A = (R,R^a)$, $\Phi_A = (\phi,\varphi_a)$, where $e^a$ and $\omega$ are the frame field and the Lorentz spin connection, and $\phi$ is the dilaton field.
Fields $\omega$ and $\phi^a$ are auxiliary  expressed in terms of $e^a$ and $\varphi$ through their equations of motion which are given by respective components of the general equations \eqref{BF_eom},
\be
\label{torsion}
\ba{l}
\rd e^a +2\,\epsilon^{ab}\, e_b \, \omega  = 0\;,
\qquad
\rd \phi +2\,\epsilon^{ab}\,e_a\, \varphi_b= 0\;.
\ea
\ee
Introducing  $g_{\mu\nu} =\eta_{ab}\, e^a_\mu e^b_\nu $ and $e = \det e_\mu^a = \sqrt{-g}$, the above  constraints can be solved as\footnote{Keeping $o(2,1)$ basis elements dimensionless one can introduce a length scale in the theory by redefining  $e_\mu^a \to e_\mu^a/\ell_{_{AdS}}$, where $\ell_{_{AdS}}$ is the AdS radius.} 
\be
\label{sol_JT}
\varphi^a = -\frac{1}{2} \epsilon^{ab}\, \partial_b \phi\;,
\qquad
\omega^a = -\frac{1}{2e} \epsilon^{\mu\nu}\partial_\mu e_\nu^a\;.  
\ee
Further, we introduce  the standard scalar curvature $\cR$ so that the 2-form curvature  $R = R_{\mu\nu} \rd x^\mu  \rd x^\nu$  is  cast into the form  $R = \half(\cR+4)\sqrt{-g}\,d^2 x$. Then, expressing the scalar curvature   in terms of the metric $\cR = \cR(g)$ we recover the second-order JT action \cite{Teitelboim:1983ux,Jackiw:1984je}
\be
\label{JT_metric}
S_{_{JT}} = \frac{\kappa}{2} \int_{\cM_2} \rd^2x \sqrt{-g} \,\phi(\cR+4) \;,
\ee
which therefore describes the coupled system of the metric field $g_{\mu\nu}$ and the dilaton field $\phi$.





\section{Colored JT gravity as $su(N,N)$ BF theory }
\label{sec:coloredJT}

In this section we formulate the colored JT gravity in the BF form and derive the spectrum of fields.  

\subsection{Matrix realization}

The gauge algebra $su(N,N)$ can be decomposed as 
\begin{equation}
\label{suNN_dec}
	su(N,N)\simeq 
	\big(\mathbb{I}_N \otimes su(1,1) \big)\oplus
	\big(su(N)\otimes \mathbb{I}_2\big)\oplus 
	\big(su(N)\otimes su(1,1)\big)\;,
\end{equation}
that makes manifest that the spectrum of gauge fields can be naturally divided in three subsets (see below). Here, $\mathbb{I}_2$ and $\mathbb{I}_N$ are respective identity elements.  In the sequel, we make use of the following matrix realization of the gauge algebra   
\be
\label{suNN}
M\in su(N,N)\;:\qquad M = \left(\ba{lc}  {\bf A}&  {\bf B} \\ {\bf B}^{\dagger}&  {\bf C} \ea \right) \;,
\ee
where ${\bf A}$, ${\bf B}$, and ${\bf C}$ are $N\times N$ matrix blocks with complex entries satisfying the anti-Hermitian conditions ${\bf A}^\dagger = -{\bf A}$, ${\bf C}^\dagger = -{\bf C}$, and the trace condition $\Tr({\bf A}+{\bf C}) = 0$, the block ${\bf B}$ is arbitrary, and $\dagger$ is the Hermitian matrix conjugation. Solving the above constraints and using real parameterization we can express an arbitrary matrix $M\in su(N,N)$ as
\be
\label{su_basis}
M = s^A (J_A\otimes\idm) +it^\alpha  ({\mathbb I}_2\otimes  \bT_\alpha)+ p^{A,\alpha}(J_A\otimes \bT_\alpha)\;,
\ee    
where $s^A$, $t^\alpha$, $p^{A,\alpha}$, with indices $A=0',0,1$ and $\alpha = 1,..., N^2-1$, are real parameters (in total, there  are $4N^2 - 1$ real parameters that equals $\dim(su(N,N)$), while  $(\mathbb{I}_2, J_A)$ are $u(1,1)$ basis elements, 
\be
\label{J_matrix}
\mathbb{I}_2 = \left(\ba{lc}  1&  0 \\ 0&  1 \ea \right),
\quad
J_{0'} = \left(\ba{lr}  0& i \\ -i&  0 \ea \right),
\quad
J_0 = \left(\ba{lr}  0& 1 \\ 1&  0 \ea \right),
\quad
J_1 = \left(\ba{lr}  i&  0 \\ 0&  -i \ea \right),
\ee
and Hermitian $N\times N$ matrices $(\idm, \bT_\alpha)$  are  $u(N)$ basis elements (for more details see, e.g. \cite{Joung:2017hsi}). The basis elements in \eqref{su_basis} are identified with matrices  $J_A\otimes\idm$, ${\mathbb I}_2\otimes  \bT_\alpha$,  $J_A\otimes \bT_\alpha$ which are Kronecker products naturally inherited from the block form \eqref{suNN}. The basis elements  are known to satisfy\footnote{It directly follows from that any complex $N\times N$  matrix ${\bf A}$ can be decomposed as ${\bf A} = {\bf B} + i\, {\bf C}$, where 
{\bf B} and {\bf C} are  Hermitian matrices. Since ${\bf A}\in {\rm Mat}(N,\mathbb{C})$ then the product of any two Hermitian matrices can be expressed via the structure constants of the matrix algebra.  } 
\be
\label{gen_relations}
J_A\,J_B=-\eta_{AB}\, \mathbb{I}_2 + \epsilon_{ABC} \,J^C\;,
\qquad
\bT_\alpha\, \bT_\beta={1\over N} \delta_{\alpha\beta} \,\idm + \left(g_{\alpha\beta}{}^\gamma+if_{\alpha\beta}{}^\gamma\right) \bT_\gamma\;,
\ee
where $g_{\alpha\beta\gamma}$ and $f_{\alpha\beta\gamma}$ are respectively totally symmetric and totally anti-symmetric structure constants.\footnote{Note that totally symmetric structure constants vanish at $N=2$. In this case, identifying ${\bf T}_\alpha \sim \sigma_{a} $ we obtain  the standard Pauli matrix relation $\sigma_a \sigma_b = \delta_{ab} I + i \epsilon_{abc} \sigma_c$ underlying the $su(2)$  commutation relations. Changing from $su(2)$ to $su(1,1)$ we see that the foregoing relation reproduces the first relation in \eqref{gen_relations}.} The commutation relations of $su(1,1)$ and $su(N)$ subalgebras read 
\be
\label{glob_sym}
[J_A,J_B]=2\,\epsilon_{ABC}\, J^C\;,
\qquad
[\bT_\alpha,\bT_\beta]= 2i\,{f_{\alpha\beta}}^\gamma \,\bT_\gamma\;.
\ee      

Following \eqref{su_basis} we denote the basis elements in the $su(N,N)$ algebra as 
\be
\label{basis}
G_A = J_A\otimes\idm\;,
\qquad
G_\alpha = \mathbb{I}_2\otimes  \bT_\alpha\;,
\qquad
G_{A,\alpha} = J_A\otimes \bT_\alpha\;.
\ee
Their commutation relations are given by\footnote{\label{Kronecker}The Kronecker product satisfies the following properties: (1) it is associative and bilinear, (2) $({\bf A}\otimes {\bf B})({\bf C}\otimes {\bf D}) = ({\bf AC})\otimes ({\bf BD})$, where the round brackets imply the standard matrix product. In particular, the trace is defined as $\tr({\bf A}\otimes {\bf B}) = \tr{\bf A}\,\tr{\bf B}$. The following commutation relation is useful in practice: 
\be
\label{acom_rel}
[{\bf A}\otimes {\bf B},{\bf C}\otimes {\bf D}] = \half[{\bf A}, {\bf C}]\otimes\{{\bf B},{\bf D}\}+\half \{{\bf A},{\bf C}\}\otimes [{\bf B},{\bf D}],
\ee
where $[\cdot, \cdot]$ and $\{\cdot, \cdot\}$ are the standard matrix (anti)commutators. } 
\be
\label{SUNN_rel}
[G_A,G_B] = 2\,\epsilon_{ABC}\,G^C\;,
\qquad
[G_\alpha,G_\beta] = 2i\,f_{\alpha\beta\gamma}\,G^\gamma\;,
\qquad
[G_A,G_\alpha] = 0\;,
\ee
\be
[G_A,G_{B,\alpha}] = 2\,\epsilon_{ABC}\,G^C_{\alpha}\;,
\qquad
[G_\alpha,G_{B,\beta}] = 2i\, f_{\alpha\beta\gamma}\,G_{B}^\gamma\;,
\ee
\be
[G_{A,\alpha},G_{B,\beta}]  = \frac{2}{N}\, \epsilon_{ABC}\,\delta_{\alpha\beta}\, G^C + 2i\,\eta_{AB}\, f_{\alpha\beta\gamma}\,G^\gamma + 2\epsilon_{ABC}\, g_{\alpha\beta\gamma}\, G_{C}^\gamma  \;.
\ee
Finally, let us list the matrix traces of bilinear combinations of the basis elements. Since 
\be
\label{tr1}
\tr \idm = N\;, 
\qquad 
\tr \mathbb{I}_2 = 2\;, 
\qquad
\tr J_A = 0\;,
\qquad
\tr \bT_\alpha = 0\;,
\ee 
then using \eqref{gen_relations} and \eqref{tr1} we obtain 
\be
\label{tr2}
\tr(J_A\,J_B) = -2\,\eta_{AB}\;, 
\qquad
\tr (\bT_\alpha\,\bT_\beta) = \delta_{\alpha\beta}\;.  
\ee
From  the footnote \bref{Kronecker} and \eqref{tr1}, \eqref{tr2}, we find non-vanishing diagonal traces of bilinear combinations of \eqref{basis} to be  
\be
\label{tr3}
\tr(G_A\,G_B) = -2N\,\eta_{AB}\;,
\qquad
\tr(G_\alpha \,G_\beta) = 2\,\delta_{\alpha\beta}\;,
\qquad
\tr(G_{A,\alpha}\,G_{B,\beta}) = -2\,\eta_{AB}\,\delta_{\alpha\beta}\;,
\ee
while all off-diagonal ones are zero,
\be
\label{tr4}
\tr(G_A\,G_\alpha) = 0\;,
\qquad
\tr(G_A\,G_{B,\beta}) = 0\;,
\qquad
\tr(G_{\alpha}\,G_{B,\beta}) = 0\;.
\ee

From the commutation relations \eqref{SUNN_rel}  it follows that the linear space of the $\mathbb{A} = su(N,N)$ algebra decomposes as
\be
\label{ABC}
\mathbb{A} = \mathbb{B}\oplus \mathbb{C}\;,
\ee 
where $\mathbb{B} = su(1,1)\oplus su(N)$ is the subalgebra, $[\mathbb{B}, \mathbb{B}] = \mathbb{B}$ consisting of two mutually commuting algebras with the basis elements $G_A$ and $G_\alpha$, and  the complimentary subspace $\mathbb{C}$ is the adjoint $\mathbb{B}$-module spanned by the basis elements $G_{A,\alpha}$, i.e. $[\mathbb{B}, \mathbb{C}]= \mathbb{C}$. Moreover, since the trace relations \eqref{tr4} define the Killing form on the algebra $\mathbb{A}$ one concludes that subspaces $\mathbb{B}$ and $\mathbb{C}$ are mutually orthogonal, $\tr(\mathbb{B}\,\mathbb{C})=0$.

\subsection{Field content} The fields are 1-form $\cA = \rd x^\mu \cA_\mu(x)$ and 0-form $\cB = \cB(x)$ taking values in the adjoint of $su(N,N)$ algebra, i.e. in matrices \eqref{suNN}. In the basis \eqref{su_basis} they can be represented as  
\be
\label{spectrum}
\ba{l}
\dps
\cA=\Omega^A G_A + iA^\alpha  G_\alpha+\,W^{A,\alpha}G_{A,\alpha}\;,
\vspace{2mm}
\\
\dps
\cB=\Phi^A G_A +iB^\alpha G_\alpha +  \Psi^{A,\alpha}G_{A,\alpha}\;.
\ea
\ee
The component expansion \eqref{spectrum} defines the spectrum of $2d$ colored JT gravity:  
\begin{itemize}
\item  $\Omega^A = \rd x^\mu\, \Omega^A_\mu(x)  $ and $\Phi^A = \Phi^A(x)$ are 1-form and $0$-form fields taking values in the adjoint of  $so(2,1)\approx su(1,1) \subset su(N,N)$ subalgebra. This color singlet subsector describes the JT dilaton gravity. 

\item  $A^\alpha = \rd x^\mu\, A^\alpha_\mu(x)  $ and $B^\alpha = B^\alpha(x)$ are 
gauge 1-form and 0-form fields, which are in the $su(N)\subset su(N,N)$ adjoint representation.

\item  $W^{A,\alpha} = \rd x^\mu\,W^{A,\alpha}_\mu(x)$ and $\Psi^{A,\alpha}= \Psi^{A,\alpha}(x)$ are 1-form and 0-form taking values in the tensor product of $su(1,1)$ and $su(N)$ adjoint representations. They form $su(N)$ adjoint multiplet of colored gravitons and dilatons.   

\end{itemize}

\noindent In total, we have the  JT graviton and dilaton, $(N^2-1)$ colored gravitons and $(N^2-1)$ colored dilatons, $su(N)$ gauge fields and  matter fields in the $su(N)$ adjoint multiplet. At $N=1$ both the colored dilaton gravity sector and $su(N)$ sector trivialize so that only the standard JT gravity is left.

\subsection{First-order action} An  action for the colored JT gravity can be naturally given in the standard  BF form 
\begin{equation}
\label{cBF}
S_{_{cJT}}[\cA, \cB]=-\frac\kappa2 \int_{\cM_2} \; \tr\cB \cF\;,
\end{equation}
where $\cM_2$ is a two-dimensional manifold, $\tr$ is defined  by \eqref{tr3}-\eqref{tr4},  and $\cF = \rd\cA + \cA  \cA$ is the 2-form curvature associated with 1-form $\cA$, the  $\kappa$ is a dimensionless coupling constant.  In the basis \eqref{spectrum} the curvature decomposes as
\be
\cF=\cR^A G_A + i\cF^\alpha  G_\alpha+\,\cZ^{A,\alpha}G_{A,\alpha}\;,
\ee
where the expansion  coefficients are given by \eqref{singlet_eom}. It follows that the action \eqref{cBF} can be cast into the component form 
\begin{equation}
\label{ccBF}
S_{_{cJT}}[\cA, \cB]=\kappa\, N  \int_{\cM_2} \; \left (\Phi^A \cR_A  +  \frac{1}{N}\, B^\alpha \cF_\alpha + 
\frac{1}N\,\Psi^{A,\alpha}\cZ_{A,\alpha}\right)\;.
\end{equation}
The action is invariant with respect to the gauge transformations 
\be
\label{gauge_0}
\delta\cA=d\varepsilon+[\cA,\varepsilon]\,,
\qquad
\delta\cB=[\cB,\varepsilon]\,,
\qquad
\ee
where following \eqref{spectrum} the 0-form gauge parameter is given by 
$\varepsilon = \varepsilon^A G_{A} +i\chi^\alpha  G_{\alpha}+\,\rho^{A,\alpha}G_{A,\alpha}$. The respective equations of motion take the form of zero-curvature and  covariant constancy conditions  
\be
\label{eom_bf_A}
\cF \equiv \rd\cA+ \cA \cA=0\;,
\qquad
\nabla_\cA \cB \equiv \rd\cB+\cA \cB-\cB \cA=0\;.
\ee
The explicit component form of  \eqref{eom_bf_A}   can be found in Appendix \bref{app:BF}. There are no local physical degrees of freedom (PdoF) in the colored gravity. However, there are finitely many global PDoF arising as integration constants in the sector of $0$-forms $\cB$. Indeed, the last equation being the covariance constancy condition is solved in terms of constant  $\cB$ taken in a fixed point $x=x_0$. Therefore, $\#$PDoF$=4N^2-1$ which is the dimension of the $su(N,N)$ adjoint.

\subsection{Action for matrix-valued fields}  Below, when considering the total action with the separated  JT gravity term we explicitly evaluate partial trace related to the basis elements $(\mathbb{I}_2, J_A)$, leaving dependence on the basis elements $(\idm, \bT_\alpha)$ implicit. To this end, it is convenient to change the notation  by singling out the $u(1,1)$ basis elements  \eqref{J_matrix} in a general $\mathbb{A}$-valued $p$-form $X$ as follows   
\be
\label{spectrum_gen}
X=\left(X_{(p)}^A J_A\right)\otimes\idm + \mathbb{I}_2\otimes {\bold Y}_{(p)}+J_{A} \otimes {\bold Z}_{(p)}^{A} \;,
\ee
where ${\bold Y}_{(p)} = iY_{(p)}^{\alpha}  \bT_\alpha$ and ${\bold Z}_{(p)}^{A} = Z_{(p)}^{A,\alpha} \bT_{\alpha}$ are matrix-valued $p$-form fields which are now subject to the reality conditions ${\bold Y}_{(p)}^\dagger = -{\bold Y}_{(p)}$ and ${\bold Z}_{(p)}^\dagger = {\bold Z}_{(p)}$.  Then, recalling the basic definition \eqref{spectrum} the BF fields can be cast into the form    
\be
\label{spectrum_new}
\ba{c}
\dps
\cA=\left(\Omega^A J_A\right)\otimes\idm + \mathbb{I}_2\otimes \bA+J_{A} \otimes \bW^{A} \;,
\vspace{2mm}
\\
\dps
\cB=\left(\Phi^A J_A\right)\otimes\idm + \mathbb{I}_2\otimes \bB+J_{A} \otimes \bPsi^{A}\;.
\ea
\ee
Using the  algebra decomposition $\mathbb{A} = \mathbb{B}+\mathbb{C}$ \eqref{ABC} we represent $\mathbb{A}$-valued fields as $X = X_{\mathbb{B}} + X_{\mathbb{C}}$, where  the  $\mathbb{B}$-valued part corresponds to the first two terms in \eqref{spectrum_gen}, while the $\mathbb{C}$-valued part is given by the third term. Then,  
\be
\label{dec_AB}
\cA = \cA_{\mathbb{B}} + \cA_{\mathbb{C}}\;,
\qquad
\cB = \cB_{\mathbb{B}} + \cB_{\mathbb{C}}\;.
\ee
Using \eqref{dec_AB} in the original BF action \eqref{cBF} one gets the factorized form of the action  
\begin{equation}
\label{cBF_dec}
S_{_{cJT}}[\cA_{\mathbb{B},\mathbb{C}}, \cB_{\mathbb{B},\mathbb{C}}]=
-\frac\kappa2 \int_{\cM_2} \; 
\tr\left[\cB_{\mathbb{B}} \cF_{\mathbb{B}}+ \cB_{\mathbb{C}}\, \nabla_{\mathbb{B}}\, \cA_{\mathbb{C}}+ \left(\cB_{\mathbb{B}}+\cB_{\mathbb{C}}\right)\cA_{\mathbb{C}} \cA_{\mathbb{C}}  \right]\;,
\end{equation}
where $\nabla_{\mathbb{B}}$ is the $\mathbb{B}$-covariant derivative defined on $\mathbb{A}$-valued  $p$-forms as follows
\be
\label{B_der}
\nabla_{\mathbb{B}} X_{(p)} = d X_{(p)} + \cA_{\mathbb{B}} X_{(p)} + (-)^{p+1} X_{(p)} \cA_{\mathbb{B}}\;.  
\ee  
The respective equations of motion   
\be
\label{BC_eom}
\frac{\delta S_{_{cJT}}}{\delta \cB_{\mathbb{B}}} = \cF_{\mathbb{B}} 
+ (\cA_{\mathbb{C}}  \cA_{\mathbb{C}})_{\mathbb{B}} = 0\;,
\qquad
\frac{\delta S_{_{cJT}}}{\delta \cB_{\mathbb{C}}} = \nabla_{\mathbb{B}} \cA_{\mathbb{C}} 
+ (\cA_{\mathbb{C}}  \cA_{\mathbb{C}})_{\mathbb{C}} = 0\;,
\ee
\be
\label{BC_eom2}
\frac{\delta S_{_{cJT}}}{\delta \cA_{\mathbb{B}}} = \nabla_{\mathbb{B}}\, \cB_{\mathbb{B}} 
+ ([\cA_{\mathbb{C}}, \cB_{\mathbb{C}}])_{\mathbb{B}}= 0\;,
\qquad
\frac{\delta S_{_{cJT}}}{\delta \cA_{\mathbb{C}}} = \nabla_{\mathbb{B}}\, \cB_{\mathbb{C}} 
+ [\cA_{\mathbb{C}},\cB_{\mathbb{B}}]
+ ([\cA_{\mathbb{C}}, \cB_{\mathbb{C}}])_{\mathbb{C}}= 0 \;,
\ee
also provides a systemic view of the component equations \eqref{singlet_eom}, \eqref{zero_form_eom}.   Taking $\mathbb{A}$-valued $p$-form in round brackets $(X_{(p)})_{\mathbb{B,C}}$ means projecting onto subspaces $\mathbb{B}, \mathbb{C}\subset \mathbb{A}$.

Equations \eqref{BC_eom}  and \eqref{BC_eom2} admit the obvious solution with all $\mathbb{C}$-valued fields set to zero. This leaves only $\mathbb{B}$-valued fields so that the resulting truncated theory is given by JT gravity  plus $su(N)$ BF theory. The subalgebra ${\mathbb B} = su(1,1)\oplus su(N)$  can be further restricted to  one of its 
parts yielding respective solutions of the original equations.
In particular, isolating the $su(1,1)\subset{\mathbb B}$ sector according to \eqref{spectrum_new} we can introduce   
\be
\label{spectrum_new_grav}
\cA=\left(\Omega^A J_A\right)\otimes\idm\;,
\qquad 
\cB=\left(\Phi^A J_A\right)\otimes\idm \;,
\ee
and, correspondingly, $(\bA=0$, $\bW^{A} =0)$ and $(\bB=0$, $\bPsi^{A}=0)$. The respective colour-singlet equations of motion reproduce the standard BF JT equations \eqref{BF_eom}.

The derivative $\nabla_{\mathbb{B}}$ acting on the $\mathbb{C}$-valued part of a $p$-form $X$ \eqref{spectrum_gen} is actually reduced to $J_{A} \otimes \nabla {\bold Z}_{(p)}^{A}$, where  
\be
\label{new_der}
\nabla {\bold Z}_{(p)}^{A} = \rd {\bold Z}_{(p)}^{A} + 2 \,\epsilon^A{}_{BC} \,\Omega^B \, {\bold Z}_{(p)}^{C} + \bA {\bold Z}_{(p)}^{A}+(-)^{p+1}{\bold Z}_{(p)}^{A}\bA \;.
\ee
Using this formula along with \eqref{spectrum_new} and partially evaluating the trace we can give an alternative   representation of the action \eqref{cBF_dec} in terms of the matrix-valued fields as 
\be
\label{action_0}
\ba{l}
\dps
S_{_{cJT}} = \kappa
\int \left( N\,\Phi^A\, R_{A}  -  \tr(\bB\, \bF) +  \tr(\bPsi^A \,\nabla\bW_A)+  \tr(\bB\, \bW_A\, \bW^A)\right.
\vspace{3mm}
\\
\hspace{20mm}+  \left.\epsilon_{ABC}\,\Phi^A\, \tr(\bW^B\, \bW^C)  +\epsilon_{ABC}\,\tr( \bPsi^A\, \bW^B\, \bW^C)\right)\;,
\ea
\ee
where the trace is now given only by the second relation in \eqref{tr2}.
Note that the colour-singlet gravity field enters  the action through the JT term plus minimal coupling to the colored fields. The interacting part contains at most cubic terms. The respective equations of motion  are given by, in the singlet sector, 
\be
\label{eq_matr1}
R^A + \frac{1}{N}\, \epsilon^{ABC}\,\tr (\bW_{B}\bW_{C})= 0\;,
\ee
\be
\label{eq_matr11}
\nabla \Phi^A +  \frac{2}{N}\, \epsilon^{ABC}\, \tr(\bW_{B}\bPsi_{C})= 0\;,
\ee
in the $su(N)$ sector,
\be
\label{eq_matr2}
\bF-\bW_A \bW^A = 0,
\qquad
\nabla \bB -[\bPsi_A, \bW^A] = 0,
\ee
in the non-singlet sector,
\begin{equation}
\label{eq_matr3}
\nabla\bW^A + \epsilon^{ABC} \left(\bW_B \bW_C - \frac1N\tr (\bW_B \bW_C)\,\idm\right) = 0\,,
\ee
\be
\label{eq_matr31}
\nabla\bPsi^A -2 \epsilon^{ABC}\Phi_B \bW_C + [\bW^A,\bB] +\epsilon^{ABC}\{\bW_B,\bPsi_C\} = 0\,.
\end{equation}
Note that the last group of terms in  \eqref{eq_matr3} is in fact the projection introduced in equations \eqref{BC_eom}-\eqref{BC_eom2}. The anticommutator in \eqref{eq_matr31} is induced from the last term in the second equation \eqref{BC_eom2} by virtue of the commutation relation given in the footnote \bref{Kronecker}.

Similar to  the representation   \eqref{spectrum_new} one can introduce the 0-form  gauge parameter 
\be
\varepsilon=\left(\varepsilon^A J_A\right)\otimes\idm + \mathbb{I}_2\otimes \bchi+J_{A} \otimes \bxi^{A}\;,
\ee
and using the relation \eqref{acom_rel} rewrite the gauge transformations \eqref{gauge_0} in the matrix basis as 
\be
\label{gauge_suNN}
\ba{l}
\dps
\delta \Omega^A =  \nabla \varepsilon^A + \frac2N \epsilon^{ABC} \tr(\bW_B \bxi_C) , 
\qquad
\dps
\delta \Phi^A = 2\epsilon^{ABC}\left(\Phi_B \varepsilon_C+ \frac1N \tr(\bPsi_B \bxi_C)\right),
\ea
\ee
\be
\ba{l}
\delta \bA =  \nabla \bchi + \{\bW^A, \bxi_A\} , 
\qquad
\delta \bB = [\bB,\bchi]+ \{\bPsi^A, \bxi_A\},
\ea
\ee
\be
\ba{c}
\dps
\delta \bW^A = \nabla \bxi^A+ [\bW^A, \bchi] + 2\epsilon^{ABC}\left(\varepsilon_B \bW_C +\half\{\bW_B, \bxi_C\}  -\frac1N\tr(\bW_B\bxi_C)\idm\right), 
\vspace{3mm}
\\
\dps
\delta \bPsi^A =[\bB, \bxi^A]+ [\bPsi^A, \bchi] +  2\epsilon^{ABC}\left(\Phi_B \bxi_C+ \varepsilon_B \bPsi_C + \half\{\bPsi_B, \bxi_C\} -\frac1N \tr(\bPsi_B\bxi_C)\idm\right).
\ea
\ee
In particular, one observes  that the singlet gravity and dilaton gauge transformations receive corrections from the matrix-valued fields.

\section{Towards the second-order  formulation of colored JT  gravity}
\label{sec:2nd_order}


In order to obtain the colored gravity action in  the second-order form  one needs to identify a set of auxiliary fields and solve the corresponding equations in terms of dynamical fields. The resulting action is given by the standard JT action \eqref{JT_metric} extended by a number of terms encoding the colored field dynamics. Following a common pattern realized in Section \bref{sec:JT} we first decompose all $o(2,1)$-valued fields in Lorentz $o(1,1)$-valued components and then identify auxiliary fields as particular Lorentz components. Fields of the model that carry $o(2,1)$ indices needed in the sequel are singlets $\Phi_A$, $\Omega^B$,  and matrices $\bPsi_A$, $\bW^B$. Their Lorentz components are denoted as 
\be
\label{aux_sing}
\Phi_A = (\phi,\uline{\varphi_a}),
\qquad
\Omega^A = (\uline{\omega},e^a),
\ee
\be
\label{aux_matr}
\bPsi_A = (\bPsi,\uline{\bPsi_a}),
\quad
\bW^A = (\uline{\bW},\bW^a),
\ee
where the underlined components are defined to be auxiliary fields: $(\omega, \varphi_a)$ in the singlet sector and $(\bW, \bPsi_a)$ in the matrix sector. The pairs  $(e^a, \phi)$,  $(\bW^a, \bPsi)$, $(\bA, \bB)$ are therefore dynamical and constitute  the colored dilaton gravity field content.\footnote{Let us note that other sets of auxiliary fields are possible which however yield  second-order formulations for different dynamical fields. In general, this leads to the so-called dual formulations of the same theory with the first-order action playing the role of a parent action. In $d$-dimensional higher-spin (linearized) gravities this effect of choosing different sets of dynamical and auxiliary fields  was investigated in  \cite{Boulanger:2003vs,Matveev:2004ac}. Our  choice of dynamical and auxiliary fields is standard and is given by the solution \eqref{sol_JT}. It is rationalized by providing the JT action as a part of the resulting colored gravity action. Dual forms of $2d$ higher-spin dynamics are also possible and some preliminary discussion can be found in  \cite{Alkalaev:2013fsa,Alkalaev:2014qpa}.}
  From the equations of motion  \eqref{eq_matr1}-\eqref{eq_matr31} represented in the Lorentz basis one finds  the auxiliary field equations in the 1-form sector [Lorentz  vector components of \eqref{eq_matr1} and \eqref{eq_matr3}]
\be
\label{full_aux_1}
\rd e^a +2\,\epsilon^{ab}\, e_b \, \omega + \frac{2}{N}\, \epsilon^{ab}\, \tr(\bW_{b}\bW) = 0\;,
\ee
\be
\label{full_aux_12}
\nabla\bW^a + \epsilon^{ab} [\bW_b, \bW] - \frac2N \epsilon^{ab} \tr(\bW_b \bW)\, \idm  = 0\;, 
\ee
and in the 0-form sector [Lorentz scalar components of \eqref{eq_matr11} and \eqref{eq_matr31}]
\be
\label{full_aux_2}
\rd \phi +2\,\epsilon^{ab}\, e_a \, \varphi_b + \frac{2}{N}\, \epsilon^{ab}\, \tr(\bW_{a}\bPsi_b) = 0\;,
\ee
\be
\label{full_aux_21}
\nabla\bPsi -2 \epsilon^{ab}\varphi_a \bW_b + [\bB,\bW] +\epsilon^{ab}\{\bW_a,\bPsi_b\} = 0.
\ee
Note that  equations \eqref{full_aux_1}, \eqref{full_aux_12} and \eqref{full_aux_2}, \eqref{full_aux_21} are obtained by variation of the action \eqref{action_0} with respect to the auxiliary 0-forms   $(\varphi_a,\bPsi_a)$ and 1-forms  $(\omega, \bW)$, respectively.  Altogether they can be traded as generalized torsion-free constraints of the colored dilaton gravity in the first-order formulation. The auxiliary equation system is linear in all auxiliary fields that allows expressing the auxiliary fields.

The auxiliary equations can be processed in two stages. Firstly, we  express the singlet auxiliary fields $(\omega, \varphi_a)$ thereby singling out the JT action. Indeed, solving the first (singlet) equations in each sector in $(\omega, \varphi_a)$ one can substitute the resulting expressions  back into the second (matrix) equations still having a  linear system with a number of equations exactly matching a number of the matrix auxiliary  field variables $(\bW, \bPsi_a)$. At the second stage, $(\bW, \bPsi_a)$ can be expressed  in terms of all dynamical fields that leads to the final second-order formulation. All in all, the auxiliary fields can be   expressed   as
\be
\ba{c}
\dps
\omega = \omega(e^a_m, \phi,\bW_b, \bPsi, \bA,\bB)\;,
\qquad
\varphi^a  = \varphi^a(e^a_m, \phi,\bW_b, \bPsi, \bA,\bB)\;,
\vspace{2mm}
\\
\dps
\bW = \bW(e^a_m, \phi,\bW_b, \bPsi, \bA,\bB)\;,
\qquad
\bPsi_a = \bPsi_a(e^a_m, \phi,\bW_b, \bPsi, \bA,\bB)\;.

\ea
\ee
These schematic expressions generalize the solutions \eqref{sol_JT} in the standard JT theory. Solving the constraints in the singlet sector yields  relations \eqref{sol_con_sing} of the next section which  provide a partial solution to the full set of constraints including the matrix sector. The full solution to the matrix constraints can be found only perturbatively.

The  action \eqref{action_0} is given in $o(2,1)$ covariant terms. As an intermediate step towards the second-order formulation it can be rewritten in terms of Lorentz $o(1,1)\subset o(2,1)$ tensor components as\footnote{It is instructive to compare this action with the 3d CS action for the (colored) gravity. It appears to be more complicated  {for} two reasons. First, contrary to a single 1-form field of CS theory,  in BF theory there two fields, a 0-form and  a 1-form. Second, $o(2,2)\approx sl(2, \mathbb{R})\oplus sl(2, \mathbb{R})$ CS theory  is naturally written in terms of Lorentz components because of the (anti)chiral factorization, while $o(2,1)\approx sl(2, \mathbb{R})$ BF theory is given in $o(2,1)$ covariant terms   that requires further decomposition in Lorentz  $o(1,1)$  components. Note also that the decomposition can be done in manifestly $o(2,1)$-covariant manner by introducing the compensator field \cite{Stelle:1979aj}.} 
\be
\label{action_lor}
\ba{c}
\dps
S_{_{cJT}} = \kappa N
\int \left( \phi\,R +  \varphi_a R^a  +  \frac1N\tr(\bB\,\bF) +  \frac1N\tr(\bPsi\,\cD\bW)
+\frac1N\tr(\bPsi^a\, \cD\bW_a)\right.
\vspace{3mm}
\\
\dps
\left. - \frac{2}{N} \epsilon_{ab} e^a\, \tr (\bPsi\, \bW^b) + \frac{2}{N} \epsilon_{ab} e^a\, \tr (\bPsi^b\, \bW) -\frac1N\tr(\bB\, \bW\, \bW)+  \frac1N\tr(\bB\, \bW_a\, \bW^a) \right.
\vspace{3mm}
\\
\dps
-  \left.\frac1N\epsilon_{ab}\,\varphi \,\tr(\bW^a\, \bW^b)+\frac{2}{N}\epsilon_{ab}\,\varphi^a \tr(\bW^b\, \bW)  
-\frac1N\epsilon_{ab}\,\tr(\bPsi\, \bW^a\, \bW^b)+\frac{2}{N}\epsilon_{ab}\,\tr (\bPsi^a\, \bW^b \,\bW)\right),
\ea
\ee
where we split off  the contribution of the $o(1,1)\oplus su(N)$ covariant derivative given by 
\be
\label{cov_der2}
\ba{c}
\cD \bZ^{a_1...a_n}_{(p)} = \rd \bZ^{a_1...a_n}_{(p)}  - 2\epsilon^{a_1}{}_b\, \omega\, \bZ^{ba_2...a_n}_{(p)} -... - 2\epsilon^{a_n}{}_b \, \omega \,\bZ^{a_n...a_{n-1}b}_{(p)}+[\bA, \bZ^{a_1...a_n}_{(p)}]
\vspace{3mm}
\\
\equiv \nabla_{_{\hspace{-1mm}L}} \bZ^{a_1...a_n}_{(p)}+[\bA, \bZ^{a_1...a_n}_{(p)}]  \;,  
\ea
\ee
with  $\nabla_{_{\hspace{-1mm}L}}$ being the Lorentz $o(1,1)$ covariant derivative.\footnote{See Appendix \bref{app:BF} for a  summary of various types of covariant derivatives introduced in this paper. }

\subsection{Generalized (singlet) torsion-free constraints} 

Let us consider first the pair of the singlet auxiliary fields $(\omega, \varphi_a)$. Their equations are given by \eqref{full_aux_1} and \eqref{full_aux_2}, 
\be
\label{wphi1}
\rd e^a +2\,\epsilon^{ab}\, e_b \, \omega + \frac{2}{N}\, \epsilon^{ab}\, \tr(\bW_{b}\bW) = 0\;,
\ee 
\be
\label{wphi2}
\rd \phi +2\,\epsilon^{ab}\, e_a \, \varphi_b + \frac{2}{N}\, \epsilon^{ab}\, \tr(\bW_{a}\bPsi_b) = 0\;,
\ee
Obviously, both equations are algebraic with respect to $\omega$ and $\varphi_a$ and  their number is sufficient (four in total) to express two components of 1-form $\omega$ and two components if 0-form $\varphi_a$ in terms of the dilaton-gravity fields $e^a, \varphi$ and the colored fields $\bW, \bW^a, \bPsi_b$. 
For future convenience we make the following  redefinition   
\be
\label{redef}
\ba{l}
\Omega^A:\;\;e^a \to e^a\;, \quad \omega \to \omega +\bar\omega\;;   
\qquad\quad\;
\Phi_A:\;\;\phi \to \phi\;, \quad\; \;\varphi_a \to \varphi_a +{\bar\varphi}_a\;,
\ea
\ee           
where new fields $\omega$ and  $\varphi_a$ satisfy the constraints with the colored graviton terms being  neglected, which therefore reproduce the constraints in the pure  BF JT theory \eqref{torsion}.\footnote{Indeed, the constraints are algebraic linear inhomogeneous equations that can be schematically represented as $\mathbb{A}x = \mathbb{B}$ provided that  $\det \mathbb{A}\neq0$ which encodes $\det e_\mu{}^a\neq0$. Splitting  $\mathbb{B} =\mathbb{B}_1 +\mathbb{B}_2$ and $x = x_1+x_2$, where $x_1$ satisfies $\mathbb{A} x_1 = \mathbb{B}_1$ we obviously find that $\mathbb{A}x_2 = \mathbb{B}_2$ and the solution is given by $x = \mathbb{A}^{-1}\mathbb{B} \equiv \mathbb{A}^{-1}\mathbb{B}_1+\mathbb{A}^{-1}\mathbb{B}_2$. In our case, $\mathbb{B}_1$ stands for $\partial e$ and $\partial \phi$, while $\mathbb{B}_2$ stands for the colored terms.} Splitting off the Lorentz connection $\omega$ is not necessary, however, it allows to manifestly control the diffeomorphism covariance. On the other hand, splitting $\varphi_a$ helps to distinguish between derivative and algebraic contributions in the interaction part of the resulting  action. Indeed, the original $\varphi_a$ enters the  action \eqref{action_lor} algebraically while the final expression  for $\varphi_a$ (see \eqref{sol_con_sing} below) is given by derivative part ($\varphi_a$) and algebraic part ($\bar \varphi_a$). Then, it follows that fields $\bar \omega$ and $\bar \varphi_a$ satisfy the following constraints  
\be
\label{con1}
\bar\omega\,  e_a \, =\, -\frac{1}{N}\, \tr(\bW_{a}\bW)\;,
\qquad \epsilon_{ab}\, e^a  \bar\varphi^b \, = \, - \frac{1}{N}\, \epsilon^{ab}\, \tr(\bW_{a}{\bf\Psi}_b)\;.
\ee
Converting the world indices of $p$-forms as $X_{\mu_1 ... \mu_p}{}^{ab...c} = e_{\mu_1}^{m_1}...e_{\mu_p}^{m_p}  X_{m_1 ... m_p|}{}^{ab...c}$
we can explicitly solve  the equation \eqref{con1} as follows 
\be
\label{opWP}
\bar\omega_m = \frac1N\, \tr(\bW_{m|b}\bW^b - \bW_{b|}{}^b\bW_m)\;,
\qquad
\bar\varphi_m = \frac1N\,\tr(\bW^{b}{}_{|m} \bPsi_b - \bW_{b|}{}^b \bPsi_m)\;.
\ee
Here, $\bW_m$ and $\bPsi_m$ are in turn the matrix auxiliary fields subject to the remaining (unsolved) constraints  \eqref{full_aux_12} and \eqref{full_aux_21}. 

To summarize, the singlet  constraints \eqref{wphi1} and  \eqref{wphi2} are  solved by 
\be
\label{sol_con_sing}
\ba{l}
\dps
\omega^m = \omega^m(\partial e)+ \frac1N\, \tr(\bW_{m|n}\bW^n - \bW_{n|}{}^n\bW_m)\;,
\vspace{3mm}
\\
\dps
\varphi^m = \varphi^m(\partial \phi) + \frac1N\,\tr(\bW^{n}{}_{|m} \bPsi_n - \bW_{n|}{}^n \bPsi_m)\;,
\ea
\ee
Here, the leading terms are given by \eqref{sol_JT} which therefore explicitly manifest the contribution of the JT sector in the full colored gravity theory. Note that the solution for the singlet auxiliary fields is linear in the matrix auxiliary fields. Keeping the Lorentz spin connection $\omega$ implicit refers to the so-called 1.5th-order formulation that allows one to simplify finding a second-order action if it derives from the original first-order action.

Using the explicit expressions for $\bar\omega$ and $\varphi_a, \bar\varphi_a$  one obtains the following form of the colored dilaton gravity action \eqref{action_lor} with partially eliminated auxiliary fields 
\be
\label{CG_action_2nd}
S_{_{cJT}} = S_{_{JT}} +  S_{_{BF}} + S_{_{X}}\;,
\ee    
where JT and $su(N)$ BF parts (see Section \bref{sec:JT}) are given by the standard expressions (modulo the overall  normalization $\kappa N$)
\be
\label{JTBF}
S_{_{JT}} = \frac{\kappa N}{2} \int_{\cM_2}  \rd^2 x \sqrt{-g} \,\phi\, (\cR+4)
\qquad \text{and} \qquad 
S_{_{BF}} = -\kappa N \int_{\cM_2} \frac1N\,\tr(\bB \bF)\;,
\ee
while all cross-terms are assembled into $S_X$ which contains  both derivative and  potential contributions

\be
\label{action_1_fin}
S_{_{X}} = \kappa N \int_{\cM_2} \rd^2 x \sqrt{-g}\, \left[\,\cK+ \cV\,\right]\;,
\ee 
where
\be
\label{der_act}
\cK =  - \frac1N\epsilon^{mn} \tr\bPsi \cD_m\bW_n
-\frac1N\epsilon^{mn}\tr\bPsi^k \cD_m\bW_{n|}{}_k 
+ \frac1N{\partial_k}  \phi\, \epsilon^{mn} \tr \bW_{m|}{}^k \bW_n\;,
\ee

\vspace{-3mm}

\be
\label{pot_term_2}
\ba{c}
\dps
\cV = \frac2N \tr(\bPsi \bW_{m|}{}^m -  \bPsi^m \bW_m) +\frac1N\phi \,\tr(\bW_{m|}{}^{m}\bW_{n|}{}^n - \bW_{m|}{}^{n}\bW_{n|}{}^m) 
\vspace{3mm}
\\
\dps
+\frac1N\epsilon^{mn}\tr\bB \big(\bW_{m|k} \bW_{n|}{}^k-\bW_m \bW_n\big) 
+\frac1N\,\tr \bPsi \big(\bW_{m|}{}^m \bW_{n|}{}^n-\bW_{m|}{}^n \bW_{n|}{}^m\big)
\vspace{3mm}
\\
\dps
-\frac{2}{N}\tr(\bPsi^m \bW_{m|}{}^n \bW_n -\bPsi^n \bW_{m|}{}^m \bW_n) 
\vspace{3mm}
\\
\dps
+\frac{2}{N^2}\tr(\bW_{m|k}\bW^k - \bW_{k|}{}^k\bW_m)\, \tr(\bPsi^m \bW_{n|}{}^n -\bPsi^n \bW_{n|}{}^m)
\vspace{3mm}
\\
\dps
-\frac{4}{N^2}\tr(\bW^{k}{}^{|m} \bPsi_k - \bW_{k|}{}^k \bPsi^m)\, \tr(\bW_{m|}{}^{n}\bW_n - \bW_{n|}{}^{n}\bW_m)\;.
\ea
\ee   
Here, all world indices in 1-forms have been converted to tangent ones  by means of the frame $e^m = e^m_\mu \rd x^\mu$ as\footnote{By a slight abuse of notation, we let $\bW_m$ denote a component of 1-form $\bW$, while a component of 1-form $\bW^a$ is denoted as $\bW_{m|}{}^a$. In what follows we distinguish between these two components simply by counting their indices.    }
\be
\label{conversion}
\bW = e^m \bW_m\;,
\qquad
\bW^a = e^m \bW_{m|}{}^a\;,
\qquad
\cD = e^m \cD_m\;.
\ee
Also, from the standard torsion-free constraint \eqref{torsion} we have $\cD e^m = 0$ that allows to represent 2-form terms as $\cD X = -e^m e^n\, \cD_m X_n$ thereby producing the action integral measure by means of  $e^m e^n = \rd^2 x \sqrt{-g}\,\epsilon^{mn}$. The base manifold ${\cM_2}$ \eqref{CG_action_2nd}-\eqref{action_1_fin} is now endowed with the metric which form is fixed by the JT equations of motion for $g_{\mu\nu}$ sourced by colored fields. Note that the potential $\cV$ contains both cubic and quartic terms given by single and double trace combinations. Apart from  the colored fields it explicitly depends on the singlet dilaton $\phi$ and  $su(N)$ BF scalar $\bB$.

\subsection{Generalized (matrix) torsion-free constraints}

The remaining matrix auxiliary fields are subject to  constraints \eqref{full_aux_12} and \eqref{full_aux_21} with the singlet auxiliary fields redefined according to \eqref{redef}: 
\be
\label{full_aux_12_1}
\cD\bW^a +2\epsilon^{ab}e_b \bW - 2\epsilon^{ab}\,\bar \omega \,\bW_b+ \epsilon^{ab} [\bW_b, \bW] - \frac2N \epsilon^{ab} \tr(\bW_b \bW)\, \idm  = 0\,, 
\ee
\be
\label{full_aux_21_2}
d\bPsi + 2 \epsilon_{ab}e^a \bPsi^b-\partial^a\phi\,\bW_a  -2 \epsilon^{ab}\bar\varphi_a \bW_b 
- [\bB,\bW] +\epsilon^{ab}\{\bW_a,\bPsi_b\} = 0\,,
\ee
where $\bar\omega$ and $\bar\varphi$ are given by \eqref{opWP}. These constraints now follow  from the action \eqref{CG_action_2nd} provided all world indices are converted to tangent ones according to \eqref{conversion}. 

Though the above constraints are  linear algebraic equations with respect to the matrix auxiliary fields they are hard to solve. However, we can still  analyze the general structure of the action \eqref{CG_action_2nd} with all auxiliary fields eliminated and derive a precise form of algebraic contributions to the  potential term. To this end, 
let us denote the matrix  auxiliary fields as 
\be
\mathbb{M}_{aux} = (\bW_m, \bPsi_n)
\ee
and the relevant parts of the  dynamical fields  as 
\be
\mathbb{D}_{mat} = (\bW_{m|k}, \bPsi)\,,
\qquad
\mathbb{G}_{scal} = (\phi, \bB)\,.
\ee
Then, the derivative \eqref{der_act} and  potential \eqref{pot_term_2} terms can be schematically represented as  
\be
\label{derivative_K}
\cK = \frac1N \mathbb{M}_{aux}(\cD \mathbb{D}_{mat}) +\frac1N\mathbb{D}_{mat}(\cD\mathbb{M}_{aux}) + 
\frac1N (\cD \mathbb{G}_{scal}) \mathbb{D}_{mat} \mathbb{M}_{aux}\;, 
\ee
\be
\label{potential_sc}
\cV = \frac1N \Big(1 +\mathbb{G}_{scal}+ \mathbb{D}_{mat}\Big)\mathbb{D}_{mat}^2  
+  \frac1N \left(1  +\mathbb{G}_{scal} +\mathbb{D}_{mat} + \frac{1}{N} \mathbb{D}_{mat}^2 \right)\,\mathbb{M}_{aux}^2 \;.
\ee       
It follows that contrary to other fields the auxiliary fields $\mathbb{M}_{aux}$ contribute to the total action \eqref{CG_action_2nd} only quadratically.  The respective equations of motion $\delta S_{_{cJT}}/\delta \mathbb{M}_{aux} =0$ can be represented in the following schematic form 
\be
\label{cons_sch}
\frac{1}{N}\left(\frac{\delta^2 \cV}{\delta \mathbb{M}_{aux}^2}(\mathbb{G},\mathbb{D})\right)^{-1}\Big(1+\mathbb{G}_{scal}\Big)\cD \mathbb{D}_{mat} \,+\, \mathbb{M}_{aux}  = 0,
\ee
which encodes the constraints \eqref{full_aux_12_1} and \eqref{full_aux_21_2} with the matrix auxiliary fields explicitly isolated. Here, we assume that the inverse operator on the left-hand side does exist in a given context. The schematic equation \eqref{cons_sch} makes it clear that substituting the solution $\mathbb{M}_{aux}$ into   $\cK+\cV$ results in reshuffling terms so that  one produces $\tilde\cK+\tilde \cV$ with $\tilde \cK$ containing second order derivative terms in the dynamical fields, 
\be
\tilde \cK = \cK +...\;\; \coloneqq\;\;\tilde\cK(\cD^2 \mathbb{D}_{mat}, \cD^2\mathbb{G}_{scal})\,,
\ee
where the dots stand for terms with derivatives coming from $\cV$, while the potential $\tilde \cV$ is obtained by setting $\mathbb{M}_{aux} =0$ in $\cV$ \eqref{potential_sc}, i.e. 
\be
\tilde \cV = \cV\big|_{_{\mathbb{F}_{aux} =0}} \;\;\coloneqq\;\; \tilde \cV(\mathbb{D}_{mat},\mathbb{G}_{scal})\,.
\ee
Thus, the residual  (at most cubic) potential is given by 
\be
\label{pot_term_3}
\ba{c}
\dps
\tilde \cV = \frac2N \tr(\bPsi \bW_{m|}{}^m) +\frac1N\phi \,\tr(\bW_{m|}{}^{m}\bW_{n|}{}^n - \bW_{m|}{}^{n}\bW_{n|}{}^m) 
\vspace{3mm}
\\
\dps
+\frac1N\epsilon^{mn}\tr\big(\bB \bW_{m|k} \bW_{n|}{}^k\big) 
+\frac1N\,\tr \bPsi \big(\bW_{m|}{}^m \bW_{n|}{}^n-\bW_{m|}{}^n \bW_{n|}{}^m\big)\;.
\ea
\ee   
Finally, the second-order action for the colored dilaton gravity can be represented in the form 
\be
S_{_{cJT}} = \frac{\kappa N}{2} \int_{\cM_2}  \rd^2 x \sqrt{-g}\, \phi (\cR+4)+\kappa  \int_{\cM_2} \tr\bB \bF +\kappa N \int_{\cM_2} \rd^2 x \sqrt{-g}\, \left[\,\tilde\cK+ \tilde\cV\,\right],
\ee
with implicit $\tilde\cK$ and explicit $\tilde \cV$ defined as above. The above form of the action manifests
that the color generalization
of JT gravity 
corresponds to the system 
consisting of the usual JT gravity,
the $SU(N)$ BF gauge theory
and the
color charged ``spin 2'' fields.
Besides the gravitational
and gauge interaction,
the latter fields have 
non-trivial self-interactions 
through the potential
$\tilde \cV$.

\section{Discussion}
\label{sec: discussions}

In this paper we have introduced a colored version of  JT gravity. It has been formulated as $su(N,N)$ BF theory. The theory describes the standard (singlet) dilaton and metric fields interacting with $su(N)$ vector and dilaton fields along with adjoint $su(N)$ multiplet of colored dilaton and metric fields. We developed both the first-order (BF) and the second-order (JT type) formulations  though explicit form of the second-order colored gravity action involves potential terms with derivatives which were obtained only implicitly.  The reason lies in complicated matrix constraints on auxiliary fields to be solved. This is a characteristic feature of the transition between  the first-order and second-order formulations which therefore allows for the perturbative consideration only.\footnote{See e.g. \cite{Vasiliev:2001wa} for the frame-like formulation of AdS$_d$ gravity and \cite{Campoleoni:2012hp} for the second-order formulation of the higher-spin Chern-Simons theory.}

Note that  the colored JT gravity is  a topological theory that is manifest in the BF form.  Local degrees of freedom can be introduced in the standard fashion by imposing appropriate boundary conditions so that a dynamics is restricted to the boundary. The respective  Schwarzian-type one-dimensional boundary theory  is considered in the paper  \cite{Alkalaev:2022qfc}.



A few comments are in order. Having in mind the isomorphisms $sl(2) \cong su(1,1) \cong o(1,2) \cong sp(2)$ one could extend the JT gauge algebra $o(1,2)$ along different classical Lie algebra series. E.g. these would be series of $sl(N)$, $su(N,M)$, $o(N,M)$, $sp(2N)$, as well as their supersymmetric and higher-spin extensions. At present, a gravity-like interpretation is available for the $sl(N)$ series in \cite{Alkalaev:2013fsa,Grumiller:2013swa,Gonzalez:2018enk} and the $su(N,N)$ series in the present paper (see Fig. \bref{fig1}). A related comment concerns a higher-spin extension of the $su(N,N)$ BF theory. Taking a formal limit $\dps\lim_{N\to \infty} su(N,N)$ one finds a higher-spin algebra $\hs$  for some particular value of $\lambda$ \cite{Bergshoeff:1989ns}.

Lastly, the colored JT gravity admits various
``colored AdS vacua'' like the colored 3d gravity
\cite{Gwak:2015vfb}. The mechanism allowing for these vacua is simple: 
we take the ansatz $\mathcal A$  
and $\mathcal B$ \eqref{spectrum_new}
with
\be
\bA=\bB=0\,,\qquad \bm W^A=
(0, e^a\,\mathbb X)\,, \qquad \bm \Psi_A=(0,\varphi_a\,\mathbb X)
\ee
for solutions of the equations \eqref{eq_matr1}-\eqref{eq_matr31}.
Here, $\mathbb X$ is a constant traceless $N\times N$ matrix belonging to $su(N)$,
and $\Omega^A=(\omega,e^a)$ and $\Phi_A=(\phi,\varphi_a)$ are the
fields of the color-singlet gravity sector.
This configuration corresponds to 
the $su(N,N)$ 1- and 0-form fields, 
\be 
\mathcal A=
\omega\,J_{0'}\otimes \mathbb I_N+e^a\,J_a\otimes \mathbb Y\,,
\qquad 
\mathcal B=
\phi\,J_{0'}\otimes\mathbb I_N+\varphi^a\,J_a\otimes \mathbb Y\,,
\label{SB bg}
\ee
where $\mathbb Y=\mathbb I_N+\mathbb X$.
The equations are 
\be 
    J_{0'}\otimes(\rd \omega\,\mathbb I_N-\epsilon^{ab}\, e_a\, e_b\,\mathbb Y^2)
    +J_a\otimes (\rd e^a+2\,\epsilon^{ab}\,e_b\,\omega)\mathbb Y=0\,,
\ee 
\be 
    J_{0'}\otimes(\rd \phi\,\mathbb I_N+2\,\epsilon^{ab}\, e_a\,\varphi_b\,\mathbb Y^2)+
    J_a\otimes (\rd \varphi^a-2\,\epsilon^{ab}\,e_b\,\phi - 2 \epsilon^{ab}\phi_b \,\omega)\mathbb Y
    =0\,.
\ee 
At this point it becomes clear that if $\mathbb Y^2=\ell^2\,
\mathbb I_N$, then we find the usual JT gravity equations in the $su(1,1)$ gauge formulation but 
with a modified cosmological constant: 
\be 
\rd \omega-\ell^2\,\epsilon^{ab}\, e_a\, e_b=0\,,
\qquad 
\rd \phi+2\,\ell^2\,\epsilon^{ab}\, e_a\, \varphi_b=0\,.
\ee 
(Here we list those JT equations which explicitly contain the cosmological constant.) The matrix $\mathbb Y$ satisfies three properties,
\be 
    \mathbb Y^\dagger =\mathbb Y\,,
    \qquad 
    \mathbb Y^2=\ell^2\,\mathbb I_N\,,
    \qquad \tr\mathbb Y=N\,.
\ee 
Using the first and second properties,
we can diagonalize $\mathbb Y$ as
\be 
    \mathbb Y=\lambda\ 
    {\rm diag}(\overbrace{-1,\ldots,-1}^{p}, 
    \overbrace{+1,\ldots, +1}^{N-p})\,,
\ee
where $p$ is the number of the negative eigenvalues.
The last condition determines $\alpha$ 
in terms of $p$ as
\be 
   \lambda^2=\left(\frac{N}{N-2p}\right)^2\,,
    \qquad p=0,1,\ldots, [\tfrac{N-1}2]\,.
\ee 
We find that the net cosmological constant 
increases in its absolute value as the  number of the color background $p$ increases.
For $p\neq 0$, the background breaks the 
color symmetry from $su(N)$ to $su(p)\oplus su(N-p)\oplus u(1)$.

Let us 
briefly discuss 
the Higgs mechanism resulting
from the symmetry breaking
at the level of the
linearized field equations
for the gauge 1-form.
For simplicity,
we choose a fluctuation 
\be 
    \eta= 
    \left(\lambda\,h^a J_{a}+h^{0'}\,J_{0'}+a\right)\otimes \mathbb S+
    \left(\lambda\,\psi^a J_{a}+\psi^{0'}\,J_{0'}+b\right)\otimes \mathbb B\,,
\ee
 associated with
a generator $\mathbb S$ of the unbroken symmetry
and a generator $\mathbb B$
of the broken symmetry satisfying
\begin{align}
 & [\mathbb Y, \mathbb S]=0\,,
    \quad 
    &\{\mathbb Y, \mathbb S\}
    =2\,\lambda\,\mathbb S\,,\\
    &[\mathbb Y,\mathbb B]=
    2\,\lambda\,\mathbb B\,,
    \quad 
     &\{\mathbb Y, \mathbb B\}=0\,.
\end{align}
The linearized flatness condition
around the background $\bar\cA$
\eqref{SB bg} reads
\be 
    \rd \eta+ [\bar\cA,\eta]=0\,,
    \label{flc eq}
\ee
and its component decomposition
 gives
\begin{alignat}{3}
     &\rd h^a-2\,\bar\omega\, \epsilon^a{}_b\,h^b
     + 2\,\epsilon^{a}{}_b\,\bar e^b\,h^{0'}=0\,,
   \qquad 
   &&\rd \psi^a-2\,\bar\omega\, \epsilon^a{}_b\,\psi^b
   + 2\,\bar e^a\,b=0\,,
   \\
    &\rd h^{0'} - 2\,\lambda^2\,
    \epsilon_{ab}\,\bar e^a\,
   h^b=0\,,
   \quad 
   && \rd \psi^{0'} =0\,,
   \\
    & \rd a =0\,,
    \quad
   &&\rd b -2\,\lambda^2\,\bar e^a\,
   \psi_a=0\,.
\end{alignat}
Note that 
the roles of the spin connection
and the spin-one mode 
are exchanged
in the symmetry broken part.
Otherwise,
$(\epsilon^{a}{}_b\,\psi^b,b)$
satisfy exactly the same equations
as those satisfied by $(h^a,a)$\,,
and hence
they also describe a 
massless spin-two mode.
This is in contrast with the 3d colored gravity \cite{Gwak:2015vfb} where the symmetry breaking generates spin-two Goldstone modes which combine with CS gauge fields to become partially-massless spin-two fields.

Let us emphasize that the basic idea behind our study is to probe various extensions of JT gravity in the BF form with a gauge algebra necessarily containing $sl(2, \mathbb{R})$ subalgebra. This latter requirement  is naturally related to that the low-energy SYK model has a conformal symmetry \cite{KitaevTalks,kitaevfirsttalk,Maldacena:2016hyu}. As is noted in the Introduction, the bulk dual of the SYK model is presently unknown and, therefore, it seems that BF theory with  extended  gauge symmetries (containing conformal sub-sector, or, equivalently, a gravitational JT sub-sector) is a natural option to investigate the dual bulk dynamics.

The color-extended gauge symmetry in our model implies that a dual SYK-like model would have the extended emergent reparametrization symmetry. Like SYK-dual of higher-spin version of JT gravity,  it will be of great interest to find a microscopic Hamiltonian for SYK-like model which has extended nearly-conformal symmetry. Furthermore, the ``colored Euclidean wormhole'' and the genus expansion of the colored JT gravity will be a tantalizing future work. This will lead us to investigate how the color extension can alter the ensemble of random matrix theory~\cite{Saad:2019lba} of the colored JT gravity and its dual SYK-like model.

In the present paper we considered the $SU(N,N)$ BF model which can be interpreted as a colored JT gravity. However, it is to be understood that in two dimensions general  gauge and matter fields can carry only scalar/spinor degrees of freedom (i.e. can be reduced to Klein-Gordon or Dirac equations) otherwise they are topological. Therefore, in the case of pure topological BF theory any direct dynamical interpretation can be ambiguous. This is true both for colored and higher-spin versions of JT gravity. The only thing  that matters here is the choice of a particular gauge algebra. 

On the other hand,  both $sl(2N)$ and $SU(N,N)$ are two real forms of the same complex algebra which means precisely that the corresponding BF theories in spite of  different space-time interpretations (see Fig. \bref{fig1}) extend JT gravity in the same conceptual direction.  Therefore, both higher-spin and colored JT gravities could be equally  relevant in finding AdS dual of SYK according to higher-spin proposals of Refs.  \cite{Gross:2017vhb,Gonzalez:2018enk,Alkalaev:2019xuv}.


\vspace{3mm}

\paragraph{Acknowledgements.} 
E.J. thanks Joaquim Gomis for his encouragement of
the current work.
The work of E.J. was supported by National Research Foundation (Korea) through the grant NRF-2019R1F1A1044065.
The work of J.Y. was supported by KIAS individual Grant PG070102 at Korea Institute for Advanced Study and the National Research Foundation of Korea (NRF) grant funded by the Korea government (MSIT) (No. 2019R1F1A1045971, 2022R1A2C1003182). J.Y. is supported by an appointment to the JRG Program at the APCTP through the Science and Technology Promotion Fund and Lottery Fund of the Korean Government. J.Y is also supported by the Korean Local Governments - Gyeongsangbuk-do Province and Pohang City.

\appendix

\section{Technical details of $su(N,N)$ BF formulation}
\label{app:BF}

\paragraph{Component form of equations.} The component form of the equations of motion in the basis \eqref{spectrum} take the following form:  

\noindent $\bullet$ 1-form sector: 
\be
\label{singlet_eom}
\ba{l}
\dps
\cR^A \equiv d \Omega^A +\epsilon^{ABC} \Omega_B  \Omega_C + \frac{1}{N}\, \epsilon^{ABC}\, W_{B,\alpha}  W_{C}^\alpha = 0\;,
\vspace{1mm}
\\
\cF_\alpha \equiv  d A_\alpha -  f_{\alpha\beta\gamma} A^\beta  A^\gamma +  f_{\alpha\beta\gamma}\, W_A{}^\beta  W^{A,\gamma} = 0\;, 
\vspace{2mm}
\\
\cZ^{A}_{\alpha} \equiv d W^{A}_{\alpha} +
\epsilon^{ABC} g_{\alpha\beta\gamma} W_{B}^\beta  W_{C}^\gamma+ \epsilon^{ABC} \Omega_B   W_{C,\alpha} - f_{\alpha\beta\gamma} A^\beta  W^{A,\gamma} = 0\;;
\ea
\ee
\noindent $\bullet$ 0-form sector: 
\be
\label{zero_form_eom}
\ba{l}
\dps
d \Phi^A +2\epsilon^{ABC} \Omega_B  \Phi_C + \frac{2}{N}\, \epsilon^{ABC}\, W_{B,\alpha}  \Psi_{C}^\alpha = 0\;,
\vspace{2mm}
\\
d B_\alpha - 2 f_{\alpha\beta\gamma} A^\beta B^\gamma + 2 f_{\alpha\beta\gamma}\, W_A^\beta  \Psi^{A,\gamma} = 0\;, 
\vspace{2mm}
\\
d \Psi^{A}_\alpha  + 2\epsilon^{ABC} \Omega_B  \Psi_{C,\alpha} 
-2 f_{\alpha\beta\gamma} A^\beta \Psi^{A,\gamma} +
\vspace{2mm}
\\
\hspace{10mm}
+2\epsilon^{ABC} g_{\alpha\beta\gamma} W_{B}^\beta \Psi_{C}^\gamma
+2f_{\alpha\beta\gamma} B^\beta W^{A, \gamma} - 2\epsilon^{ABC} \Phi_B W_{C,\alpha}= 0\;. 
\ea
\ee
Using \eqref{gen_relations} one can show that these equations are equivalent to the matrix equations \eqref{eq_matr1}-\eqref{eq_matr3}. Note that the  equations of motions of pure JT and $su(N)$ BF theories are now sourced by terms with colored fields.

\paragraph{Covariant derivatives.} Throughout the paper we introduced a number of covariant derivatives associated to various subalgebras of the original gauge algebra $\mathbb{A} =su(N,N)$. It is useful to bring them together: 
\begin{itemize}
    \item $\nabla_{\cA}$ is a covariant derivative \eqref{eom_bf_A}; 
    \vspace{-2mm}
    \item $D$ is a $o(1,2)$ covariant derivative;
    \vspace{-2mm}
    \item $\nabla_{\mathbb{B}}$ is a covariant derivative \eqref{B_der} of the subalgebra $\mathbb{B}\subset\mathbb{A}$ \eqref{ABC};
    \vspace{-2mm}
    \item $\nabla$ is a tensor product part of $\nabla_{\mathbb{B}}$ acting on $\mathbb{C}$-valued fields \eqref{new_der};
    \vspace{-2mm}
    \item $\cD$ is $o(1,1)\oplus su(N)$ covariant derivative \eqref{cov_der2};
    \vspace{-2mm}
    \item $\nabla_L$ is the Lorentz $o(1,1)$ covariant derivative being a part of $\cD$ \eqref{cov_der2}. 
\end{itemize}

\bibliographystyle{JHEP}
\bibliography{colorJT}

\end{document}